\documentclass[
  aps,
  pre,
  reprint,
  superscriptaddress,
  longbibliography,
  nofootinbib,
  floatfix
]{revtex4-2}

\usepackage{amsmath,amssymb}
\usepackage{graphicx}
\usepackage{bm}
\newcommand{\alpham}{\alpha_{\mathrm{m}}}

\begin{document}

\title{Dynamical crossover from motor-dominated to drag-dominated transport in a minimal active transport network}

\author{Kazuhiko Mitsuhashi}
\affiliation{Department of Electrical and Electronic Engineering, National Institute of Technology, Sasebo College, Sasebo, Japan}

\date{July 6, 2026}

\begin{abstract}
Motor-driven intracellular transport is often described in terms of motor activity, but macroscopic transport also depends on how effectively motor-generated force is converted into coherent motion. Motivated by cytoplasmic streaming, a minimal active transport network is examined in which motor-driven transport competes with an effective slip-related dissipative resistance. The model is not intended as a quantitative reconstruction of \textit{Nitella} cytoplasmic streaming, but as a minimal system for isolating the relation between motor activity, resistance, and transport output.

A controlled scan over $\gamma_{\mathrm{Slip}}$ and $\alpham$, with three independent seeds per condition, shows that increasing $\gamma_{\mathrm{Slip}}$ strongly suppresses mean transport speed while leaving the motor-bound fraction nearly unchanged. The mean load and motor force remain finite in the high-$\gamma_{\mathrm{Slip}}$ regime, indicating that motors remain mechanically active even when transport is suppressed. The dependence of transport speed on $\alpham$ progressively disappears with increasing $\gamma_{\mathrm{Slip}}$: the motor dominance ratio decreases from $R\approx1.69$ to $R\approx1.01$, and the corresponding velocity difference decreases from $\sim1.9~\mu\mathrm{m/s}$ to $\sim0.003~\mu\mathrm{m/s}$.

These results indicate a dynamical crossover from motor-dominated to drag-dominated transport. The minimal model provides a compact physical scenario in which active force generation persists while its contribution to net transport is suppressed by increased effective dissipative resistance.
\end{abstract}

\maketitle

\section{Introduction}

Motor-driven intracellular transport is a fundamental mechanism by which living cells organize and redistribute intracellular material. In many systems, molecular motors convert chemical energy into mechanical work and generate directed motion along cytoskeletal tracks~\cite{Howard2001,ValeMilligan2000,Julicher1997}. However, the macroscopic transport output of an active system is not determined by motor activity alone. The conversion of microscopic motor force into coherent transport also depends on mechanical resistance, collective coupling, and effective dissipation. Thus, a central physical question is how motor activity is converted into macroscopic transport, and under what conditions this conversion becomes inefficient.

Cytoplasmic streaming provides a striking example of large-scale intracellular transport driven by active processes. In characean cells such as \textit{Chara} and \textit{Nitella}, the streaming of endoplasm has long served as a classical system for studying the relation between cytoplasmic organization, active force generation, and cellular-scale flow. The experimental basis of this field was established through careful studies of giant algal cells, including historical and methodological work by Kamiya, measurements of the velocity distribution of protoplasmic streaming in \textit{Nitella} cells by Kamiya and Kuroda, analyses of the motive force driving streaming by Tazawa, and cell-model experiments by Shimmen and Tazawa~\cite{Kamiya1986,KamiyaKuroda1956,Tazawa1968,ShimmenTazawa1982}. These studies provide an important foundation for considering cytoplasmic streaming as a physical transport problem.

Later work clarified the molecular and cellular basis of plant cytoplasmic streaming, including the roles of actin organization, myosin motors, organelle transport, and biochemical regulation~\cite{ShimmenYokota2004,Yokota2003,Tominaga2013}. In parallel, physical models have examined hydrodynamic transport, wall slip, and microfilament organization in streaming cells~\cite{GoldsteinMeent2015,Meent2008,Wolff2012,WoodhouseGoldstein2013}. These studies show that cytoplasmic streaming is not determined by motor activity alone, but by the coupling between active force generation, cytoplasmic organization, and mechanical resistance. The present study follows this minimal-modeling direction, while deliberately avoiding a quantitative reconstruction of the full biological system.

Here, a minimal modeling approach is taken. Rather than attempting to reproduce \textit{Nitella} cytoplasmic streaming quantitatively, a cytoplasmic-streaming-inspired active transport network is constructed to ask how its transport response changes when effective dissipative resistance is varied. The model contains motor-driven active transport, controlled by the dimensionless motor-activity parameter $\alpham$, and an effective slip-related dissipative resistance, controlled by $\gamma_{\mathrm{Slip}}$. This formulation is related to broader active-matter descriptions in which microscopic force generation and macroscopic mechanical response need not be trivially proportional~\cite{Ramaswamy2010,Marchetti2013}.

The central question of this work is whether an active transport network can remain mechanically active while losing its dependence on motor activity at the level of macroscopic transport. To address this question, a controlled $\gamma_{\mathrm{Slip}}$ scan was performed over three values of $\alpham$. Increasing $\gamma_{\mathrm{Slip}}$ strongly suppresses mean transport speed, while the motor-bound fraction remains nearly unchanged and the motor force remains finite. Moreover, the dependence of transport speed on $\alpham$ progressively disappears as $\gamma_{\mathrm{Slip}}$ increases. These results indicate a dynamical crossover from motor-dominated to drag-dominated transport.

This study therefore provides a compact physical scenario in which motor activity and macroscopic transport output become decoupled through increased effective dissipative resistance. The aim is not to identify a critical transition or a universal scaling law, but to establish a minimal and reproducible example of how active force generation can persist while its contribution to net transport is suppressed.

\section{Methods}

\subsection{Minimal active transport network model}
\label{sec:model}

A minimal active transport network model was constructed, inspired by cytoplasmic streaming but not intended as a quantitative reconstruction of \textit{Nitella} cytoplasm. The model was designed to examine how active motor-driven transport is modified by effective dissipative resistance. In this first analysis, the focus is on the emergence of a dynamical crossover between motor-dominated and drag-dominated transport.

The system consisted of mobile particles in a three-dimensional rectangular domain. A subset of the particles carried motor-like activity. Motor-bearing particles could bind near the cortex-side boundary, where an actin-like attachment region was represented by periodic stripes in the transverse direction. When bound, motor-bearing particles generated directed motion along the longitudinal $x$ direction. This construction is intentionally minimal: the model does not explicitly resolve individual actin filaments, myosin molecules, organelles, or hydrodynamic flow fields.

The effective level of motor activity was controlled by the dimensionless parameter $\alpham$, while the effective dissipative resistance near the cortex-side boundary was controlled by $\gamma_{\mathrm{Slip}}$. At the coarse-grained level, the particle dynamics were represented as an overdamped force balance,
\begin{equation}
\gamma_i(z_i)\dot{\mathbf{r}}_i
=
\mathbf{F}^{\mathrm{int}}_i
+
\mathbf{F}^{\mathrm{wall}}_i
+
\mathbf{F}^{\mathrm{act}}_i,
\label{eq:eom}
\end{equation}
where $\mathbf{F}^{\mathrm{wall}}_i$ denotes boundary forces and $\mathbf{F}^{\mathrm{act}}_i$ denotes motor-driven active force. Here, $\mathbf{F}^{\mathrm{int}}_i$ denotes effective internal or constraint forces used to represent attachment-plane constraints and other coarse-grained mechanical restrictions. It should not be interpreted as a force derived from a prescribed graph of permanent interparticle bonds.

Because the word ``network'' can imply a prescribed graph or a cross-linked polymer structure, we specify its meaning in the present model. Here, a network means a coarse-grained active transport medium formed by mobile particles, motor-like active particles, cortex-side attachment stripes, local drag, and boundary or attachment constraints. It is not a fixed graph: no permanent particle--particle edges, FENE bonds, bond-rupture rules, or molecularly resolved actin, endoplasmic-reticulum, or organelle network are included in the present paper-facing implementation. The relevant coupling is the collective mechanical response of the particle assembly to motor forcing and slip-related resistance.

The local drag coefficient in the cortex-side slip layer was controlled by
\begin{equation}
\gamma_i(z_i)
=
\gamma_{\mathrm{Slip}}
+
\left(\gamma_{\mathrm{unit}}-\gamma_{\mathrm{Slip}}\right)
\min\left(1,\frac{z_i}{\ell_{\mathrm{Slip}}}\right),
\label{eq:slipdrag}
\end{equation}
where $z_i$ is the distance from the lower boundary and $\ell_{\mathrm{Slip}}=0.5~\mu\mathrm{m}$. For attached motor-bearing particles, the motor activity parameter entered both the target motor speed and the attachment attempt probability in an effective manner,
\begin{equation}
\begin{aligned}
v_i^{\mathrm{tar}}
&=
\alpham V_{\mathrm{free}}
\max\left[0,1-\frac{f_i^{\mathrm{load}}}{n_iF_{\mathrm{stall}}}\right],\\
F_i^{\mathrm{act}}
&=
\min\left[\gamma_i(z_i)v_i^{\mathrm{tar}},\, n_iF_{\mathrm{stall}}\right],\\
p_i^{\mathrm{att}}
&=
\min\left(1,\alpham p_0^{\mathrm{att}}\right),
\end{aligned}
\label{eq:activeforce}
\end{equation}
with $\mathbf{F}^{\mathrm{act}}_i=F^{\mathrm{act}}_i\hat{\mathbf{x}}$ for attached particles and $\mathbf{F}^{\mathrm{act}}_i=0$ otherwise. Here, the opposing load was defined operationally as
\begin{equation*}
f_i^{\mathrm{load}}
=
\max\left[0,-\left(\mathbf{F}^{\mathrm{int}}_i+\mathbf{F}^{\mathrm{wall}}_i\right)\cdot\hat{\mathbf{x}}\right],
\end{equation*}
$n_i$ is the number of bound motor units, and $p_0^{\mathrm{att}}$ is the baseline attachment-attempt probability. Equations~\eqref{eq:eom}--\eqref{eq:activeforce} define the coarse-grained force-balance structure of the model: $\alpham$ changes the effective motor activity, whereas $\gamma_{\mathrm{Slip}}$ changes the local dissipative resistance. Further implementation details are described below.

\subsection{Simulation domain, boundaries, and numerical implementation}

Simulations were performed in a rectangular three-dimensional domain with dimensions
\begin{equation*}
L_X=50~\mu\mathrm{m}, \quad
L_Y=30~\mu\mathrm{m}, \quad
L_Z=30~\mu\mathrm{m}.
\end{equation*}
Each run contained
\begin{equation*}
N_{\mathrm{TOTAL}}=12000
\end{equation*}
mobile particles. At initialization, particles were distributed randomly in the domain, with positions periodic in the $x$ and $y$ directions and bounded in the $z$ direction. The lower boundary was located at $z=0$, and the upper tonoplast-side boundary was represented near $z=28~\mu\mathrm{m}$. Wall repulsion was applied in the $z$ direction to keep particles within the physical domain.

The cortex-side attachment region was located near the lower boundary. The actin-like layer thickness was
\begin{equation*}
0.15~\mu\mathrm{m},
\end{equation*}
and the bound-plane offset from the lower boundary was
\begin{equation*}
0.02~\mu\mathrm{m}.
\end{equation*}
Attachment was allowed only for motor-bearing particles located near the bottom region and within periodic transverse stripes. Bound particles were constrained toward the cortex-side bound plane while generating directed motion along the $x$ direction.

Consistent with Eq.~\eqref{eq:eom}, the local drag force was proportional to particle velocity. Near the lower boundary, the local friction coefficient followed Eq.~\eqref{eq:slipdrag} within a slip layer of thickness
\begin{equation*}
0.5~\mu\mathrm{m}.
\end{equation*}
Away from this layer, the friction approached the baseline value
\begin{equation*}
\gamma_{\mathrm{unit}}=0.024~\mathrm{pN\,s/\mu m}.
\end{equation*}
An additional tangential drag was applied near the tonoplast-side boundary to suppress unrealistically fast tangential motion near the upper boundary.

The simulations used a time step
\begin{equation*}
\Delta t=0.002~\mathrm{s}
\end{equation*}
with five substeps per output frame. Each run was simulated until $t=30~\mathrm{s}$. For the paper-facing analysis, the late-time interval from $t=20$ to $30~\mathrm{s}$ was used, after the initial transient behavior had decayed. The same analysis window was used for the mean transport velocity, motor-bound fraction, load, and motor force. This choice was applied uniformly to all runs in the main parameter scan.

The simulations were implemented in Python using Taichi with CUDA acceleration~\cite{Hu2019} on a five-node GPU workstation cluster. The computational workflow used a manifest-based batch execution system. Each run was assigned a unique run identifier, and all simulation parameters and output files were saved in a dedicated run directory. The raw outputs used in this paper were taken from the main $\gamma_{\mathrm{Slip}}$ parameter scan directory.

\subsection{Motor activity and dissipative resistance}

Motor-bearing particles could attach within the cortex-side attachment region and generate directed motion along the longitudinal $x$ direction. Once attached, a particle was assigned an integer number of bound motor units $n_i$, sampled uniformly from $\{1,2,3\}$. The motor-driven target velocity decreased with opposing load and was limited by an effective stall-force scale, as summarized in Eq.~\eqref{eq:activeforce}. The resulting directed motor force was computed from the local drag balance and clipped by the available stall force.

The baseline free motor speed was
\begin{equation*}
V_{\mathrm{free}}=60~\mu\mathrm{m/s},
\end{equation*}
and the single-unit stall-force scale was
\begin{equation*}
F_{\mathrm{stall}}=0.7~\mathrm{pN}.
\end{equation*}
The baseline attachment-attempt probability was
\begin{equation*}
p_0^{\mathrm{att}}=1.0
\end{equation*}
per eligible attachment attempt. These values are treated as effective model parameters. They are not intended as direct quantitative measurements of \textit{Nitella} molecular motors.

The parameter $\alpham$ was used as an effective motor-activity control. In the implementation used here, it modifies both motor-driven motion and the probability of motor attachment attempts. Thus, $\alpham$ should be interpreted as a coarse-grained activity parameter, not as a direct scaling of a single molecular quantity. Note that, because $p_0^{\mathrm{att}}=1.0$, the attachment-probability channel in Eq.~\eqref{eq:activeforce} saturates for $\alpham\geq 1.0$; the $\alpham$-dependence of transport between $\alpham=1.0$ and $2.0$ therefore arises primarily through the target-velocity term.

The parameter $\gamma_{\mathrm{Slip}}$ was used as an effective dissipative-resistance control. In the implementation used here, it sets the local drag scale assigned to the cortex-side slip layer, while the friction away from this layer approaches the baseline value. Within the present scan, increasing $\gamma_{\mathrm{Slip}}$ therefore raises the local resistance opposing the conversion of motor activity into net transport. This formulation allows motor binding and motor force generation to remain present while the resulting macroscopic transport is suppressed.

\subsection{Main slip-resistance parameter scan}

The main dataset analyzed in this paper was a controlled $\gamma_{\mathrm{Slip}}$ parameter scan. This scan was designed to test how increasing effective dissipative resistance changes the dependence of transport on motor activity.

Eight values of $\gamma_{\mathrm{Slip}}$ were examined:
\begin{equation*}
0.005,\ 0.010,\ 0.015,\ 0.020,\ 0.030,\ 0.050,\ 0.080,\ 0.100.
\end{equation*}
For each value of $\gamma_{\mathrm{Slip}}$, three values of $\alpham$ were simulated:
\begin{equation*}
0.5,\ 1.0,\ 2.0.
\end{equation*}
For each $\gamma_{\mathrm{Slip}}$--$\alpham$ condition, three independent seeds were used. The complete dataset therefore consisted of
\begin{equation*}
8 \times 3 \times 3 = 72
\end{equation*}
independent simulation runs. All 72 runs were included in the paper-facing analysis. Earlier exploratory calculations were used only for parameter selection and stability checks, and are not used as primary evidence in the main figures.

\subsection{Measured quantities}

For each run, the following quantities were evaluated over the late-time interval $t=20$ to $30~\mathrm{s}$.

The mean transport velocity was measured as the simulation-reported mean longitudinal transport velocity,
\begin{equation*}
\langle V_x \rangle .
\end{equation*}
Because the sign of the velocity reflects the chosen coordinate convention, transport speed was reported as
\begin{equation*}
|\langle V_x \rangle|.
\end{equation*}

The motor-bound fraction was measured as the fraction of motor-bearing particles in the bound state during the analysis interval. This quantity was used to distinguish transport suppression caused by reduced motor binding from transport suppression caused by reduced force-to-motion conversion.

The mean load was measured as the average load opposing motor-driven motion among bound particles. The mean motor force was measured as the average directed motor force among bound particles over the same interval. These quantities were used to determine whether motors remained mechanically active in the high-$\gamma_{\mathrm{Slip}}$ regime.

For each $\gamma_{\mathrm{Slip}}$--$\alpham$ condition, the reported values are the mean over three independent runs. Error bars in the figures indicate the standard deviation across these runs. Because the number of seeds is limited to three, the error bars are used to summarize run-to-run variability rather than to support formal significance testing. The crossover is therefore presented as a systematic trend across the eight sampled $\gamma_{\mathrm{Slip}}$ conditions, rather than as a claim resting on formal significance testing at any single condition.

\subsection{Crossover metrics and data processing}
\label{sec:crossover-metrics}

To quantify the dependence of transport on motor activity, the low- and high-$\alpham$ conditions were compared at each value of $\gamma_{\mathrm{Slip}}$. The motor dominance ratio was defined as
\begin{equation}
R =
\frac{|V_{\alpham=2.0}|}
{|V_{\alpham=0.5}|}.
\end{equation}
When $R>1$, transport speed depends on $\alpham$, indicating a motor-dominated regime. When $R$ approaches unity, the transport speed becomes nearly independent of $\alpham$, indicating a drag-dominated regime in which motor activity remains present but no longer controls the net transport speed.

The $\alpham$-dependent velocity difference was also defined as
\begin{equation}
\Delta V =
|V_{\alpham=2.0}|
-
|V_{\alpham=0.5}|.
\end{equation}
A decrease of $\Delta V$ toward zero indicates loss of $\alpham$ dependence. These two metrics were used together to identify the dynamical crossover from motor-dominated to drag-dominated transport.

The crossover scale was treated as operational rather than critical. Two simple criteria were used in the summary analysis: the first $\gamma_{\mathrm{Slip}}$ value at which $R \leq 1.05$, and a log-interpolated estimate of the $\gamma_{\mathrm{Slip}}$ value at which $R=1.1$. These values were used only to summarize the crossover scale and were not interpreted as evidence of a phase transition or universal critical behavior.

All paper-facing data processing was performed using a dedicated analysis pipeline. First, the raw run directories were checked for the presence of parameter files, summary files, and time-series outputs. Second, each run was converted into a tidy table containing the run identifier, host, seed, $\gamma_{\mathrm{Slip}}$, $\alpham$, mean transport velocity, motor-bound fraction, mean load, and mean motor force. Third, runs with the same $\gamma_{\mathrm{Slip}}$ and $\alpham$ were averaged across the three independent seeds. Finally, the crossover metrics $R$ and $\Delta V$ were computed from the seed-averaged transport speeds.

The main figures were generated from the processed tables. The seed-averaged $\gamma_{\mathrm{Slip}}$--$\alpham$ table was used for Figs.~\ref{fig:speed_bound} and~\ref{fig:load_motorF}, and the crossover-metric table was used for Fig.~\ref{fig:crossover}. Figures were exported in PNG, PDF, and SVG formats. The PDF versions were used as the primary publication-quality figure files.

\subsection{Scope of interpretation}

The model and analysis were intentionally kept minimal (Sec.~\ref{sec:model}), and the crossover metrics are treated as operational rather than critical (Sec.~\ref{sec:crossover-metrics}). Within that scope, the purpose of the main parameter scan was to test whether a simple active transport network can exhibit a dynamical crossover from motor-dominated to drag-dominated transport. The analysis below establishes that $\alpham$ dependence is strong at low $\gamma_{\mathrm{Slip}}$ and nearly disappears at high $\gamma_{\mathrm{Slip}}$.

\section{Results}

\subsection{Main parameter scan}

The main $\gamma_{\mathrm{Slip}}$ scan consisted of 72 simulations, covering eight values of $\gamma_{\mathrm{Slip}}$, three values of $\alpham$, and three independent seeds for each parameter condition. The analysis below uses seed-averaged quantities for each $\gamma_{\mathrm{Slip}}$--$\alpham$ condition.

\subsection{Flow suppression without loss of motor binding}

\begin{figure}[!htbp]
\centering
\includegraphics[width=0.98\columnwidth]{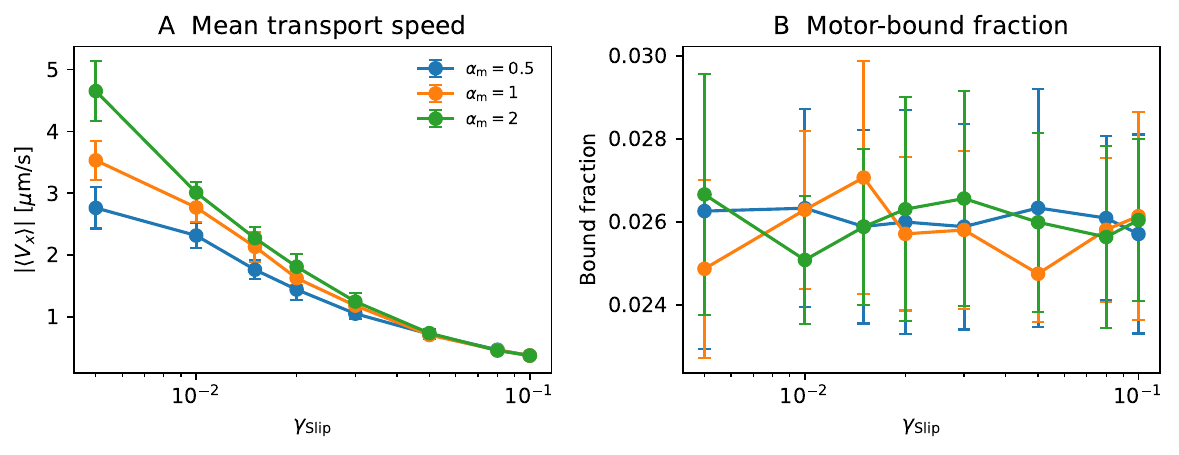}
\caption{Flow suppression without loss of motor binding. (A) Mean transport speed, $|\langle V_x\rangle|$, as a function of $\gamma_{\mathrm{Slip}}$ for three values of $\alpham$. (B) Motor-bound fraction for the same conditions. Points are means over three independent runs; error bars indicate standard deviations.}
\label{fig:speed_bound}
\end{figure}

Increasing $\gamma_{\mathrm{Slip}}$ strongly suppressed the mean transport speed in all $\alpham$ conditions (Fig.~\ref{fig:speed_bound}A). The mean absolute longitudinal velocity, $|\langle V_x\rangle|$, decreased across the scanned range, showing that increased slip-related resistance limits net transport in the minimal active transport network.

By contrast, the motor-bound fraction remained nearly unchanged (Fig.~\ref{fig:speed_bound}B). Across all $\gamma_{\mathrm{Slip}}$--$\alpham$ conditions, the mean bound fraction was $0.02596$, with a range from $0.02475$ to $0.02707$ and a coefficient of variation of $0.0204$. Thus, transport suppression is not primarily explained by loss of motor binding. Instead, motors remain bound at a comparable level while the conversion of motor activity into net transport is reduced.

\subsection{Load and motor-force response}

\begin{figure}[!htbp]
\centering
\includegraphics[width=0.98\columnwidth]{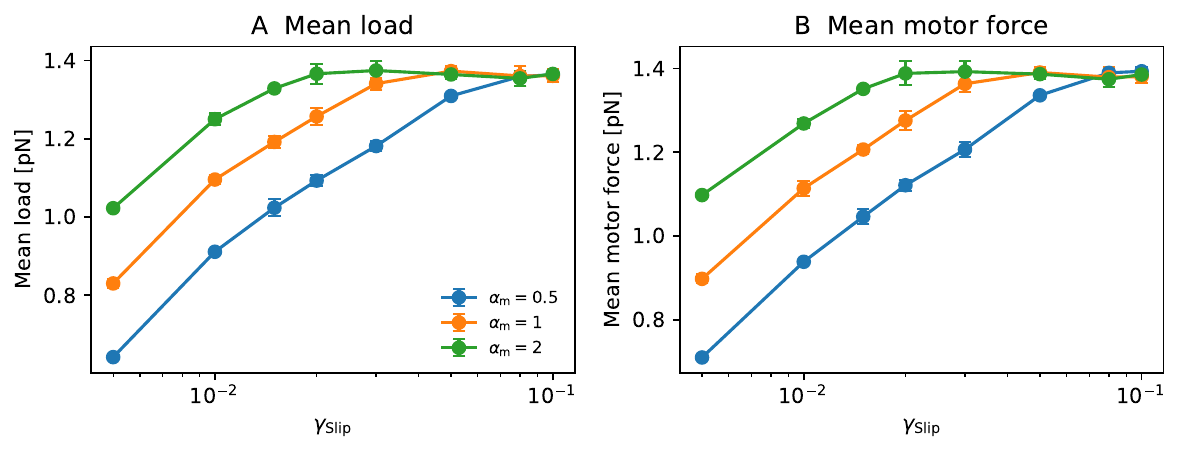}
\caption{Load and motor-force response during transport suppression. (A) Mean load and (B) mean motor force as functions of $\gamma_{\mathrm{Slip}}$ for three values of $\alpham$. Points are means over three independent runs; error bars indicate standard deviations.}
\label{fig:load_motorF}
\end{figure}

The mean load increased with $\gamma_{\mathrm{Slip}}$ and approached a saturated range at high $\gamma_{\mathrm{Slip}}$ (Fig.~\ref{fig:load_motorF}A). Across the full scan, the mean load ranged from $0.642$ to $1.37$. The mean motor force showed a similar response, increasing from $0.710$ to $1.39$ and remaining finite in the high-$\gamma_{\mathrm{Slip}}$ regime (Fig.~\ref{fig:load_motorF}B).

These results indicate that transport suppression is not caused by a loss of mechanical motor activity. Motors remain bound and force-generating, but their ability to produce net transport is increasingly limited by effective dissipative resistance.

\subsection{Dynamical crossover from motor-dominated to drag-dominated transport}

\begin{figure}[!htbp]
\centering
\includegraphics[width=0.98\columnwidth]{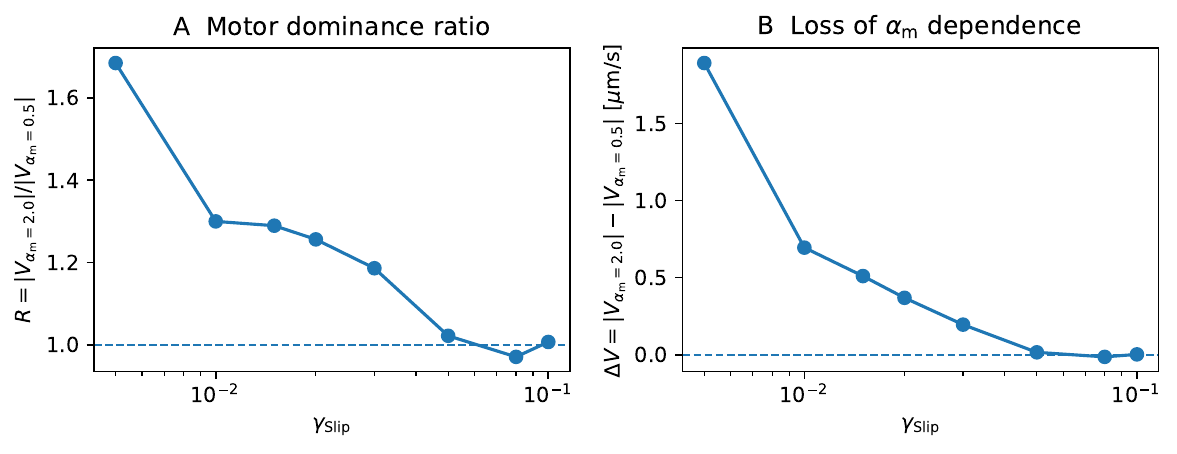}
\caption{Dynamical crossover from motor-dominated to drag-dominated transport. (A) Motor dominance ratio, $R=|V_{\alpham=2.0}|/|V_{\alpham=0.5}|$, as a function of $\gamma_{\mathrm{Slip}}$. (B) Velocity difference, $\Delta V=|V_{\alpham=2.0}|-|V_{\alpham=0.5}|$. Points are computed from seed-averaged transport speeds.}
\label{fig:crossover}
\end{figure}

The dependence of transport on motor activity was quantified by the motor dominance ratio,
\begin{equation*}
R =
\frac{|V_{\alpham=2.0}|}
{|V_{\alpham=0.5}|}.
\end{equation*}
At $\gamma_{\mathrm{Slip}}=0.005$, $R\approx1.69$, indicating that transport is strongly $\alpham$-dependent in the low-resistance regime (Fig.~\ref{fig:crossover}A). The corresponding velocity difference,
\begin{equation*}
\Delta V =
|V_{\alpham=2.0}|
-
|V_{\alpham=0.5}|,
\end{equation*}
was $\sim1.9~\mu\mathrm{m/s}$ (Fig.~\ref{fig:crossover}B).

As $\gamma_{\mathrm{Slip}}$ increased, $\alpham$ dependence progressively disappeared. At $\gamma_{\mathrm{Slip}}=0.100$, $R\approx1.01$, and $\Delta V$ decreased to $\sim0.003~\mu\mathrm{m/s}$. Thus, net transport becomes nearly independent of $\alpham$ in the high-$\gamma_{\mathrm{Slip}}$ regime.

As an operational estimate (Sec.~\ref{sec:crossover-metrics}), the condition $R \leq 1.05$ was first reached at $\gamma_{\mathrm{Slip}}=0.050$, while log-interpolation to $R=1.1$ gave $\gamma_{\mathrm{Slip}}\approx 0.039$. Taken together, the results establish a dynamical crossover from a motor-dominated regime, in which transport is sensitive to motor activity, to a drag-dominated regime, in which motor activity persists but no longer controls the net transport speed.

\section{Discussion}

\subsection{Main finding}

This study examined a minimal active transport network in which motor-driven transport competes with an effective dissipative resistance. The central finding is that motor binding and motor-force generation persist even as increasing $\gamma_{\mathrm{Slip}}$ suppresses $\alpham$-dependent transport. The observed change therefore reflects how motor activity is converted into net transport, not a loss of motor activity itself.

\subsection{Suppression of force-to-motion conversion}

The main parameter scan suggests a useful distinction between motor activity and transport output. In the low-$\gamma_{\mathrm{Slip}}$ regime, increasing $\alpham$ produces a clear increase in transport speed, indicating that the transport response is motor-dominated. In the high-$\gamma_{\mathrm{Slip}}$ regime, motors still bind and generate force, but changes in $\alpham$ have little effect on the resulting transport speed. Thus, motor activity persists, but its macroscopic expression as directed transport is suppressed.

This behavior can be interpreted as a reduction in force-to-motion conversion. The active elements continue to exert force through Eq.~\eqref{eq:activeforce}, but the increased local drag described by Eq.~\eqref{eq:slipdrag} limits the ability of that force to produce coherent net motion. In this sense, the high-resistance regime should not be interpreted as an inactive limit. It is an active regime in which the transport response is controlled primarily by resistance rather than by the level of motor activity. This interpretation is consistent with broader active-matter viewpoints in which the relation between microscopic force generation and macroscopic response depends on coupling and dissipation~\cite{Ramaswamy2010,Marchetti2013}.

\subsection{Nature of the crossover}

The relevant physical content of this result is the progressive loss of $\alpham$ dependence with increasing $\gamma_{\mathrm{Slip}}$, reproduced consistently across the scanned range. Consistent with the operational definition of the crossover scale (Sec.~\ref{sec:crossover-metrics}), no attempt is made to assign it the status of a unique critical point.

For the same reason, no universal scaling exponent is assigned to the decrease of the $\alpham$-dependent velocity difference. Although the decrease of $\Delta V$ is systematic over the scanned range, the present dataset was designed to identify a crossover, not to establish asymptotic scaling behavior. The robust conclusion is that motor activity controls transport at low dissipative resistance, whereas this control is lost at high dissipative resistance.

\subsection{Implications for cytoplasmic-streaming-inspired modeling}

The model was inspired by cytoplasmic streaming, but its role is to isolate a physical mechanism rather than to reproduce a specific biological system in detail. Classical studies of characean cytoplasmic streaming established that the phenomenon is experimentally rich and mechanically subtle, involving not only visible particle motion but also the organization and mechanical state of the endoplasm~\cite{Kamiya1986,KamiyaKuroda1956,Tazawa1968,ShimmenTazawa1982}. Modern molecular and physical studies have clarified important aspects of actomyosin-driven transport, hydrodynamic flow, and streaming organization~\cite{ShimmenYokota2004,Yokota2003,Tominaga2013,GoldsteinMeent2015,Meent2008,Wolff2012,WoodhouseGoldstein2013}, but any minimal model necessarily simplifies many biological details.

The present model should therefore be viewed as complementary to, rather than a replacement for, detailed experimental and cell-model studies. It identifies a simple scenario in which motor activity and macroscopic transport output can become decoupled through increased effective dissipative resistance. In this scenario, a reduction in observed streaming speed should not, by itself, be interpreted as direct evidence for a proportional loss of motor activity.

This distinction is also relevant for future comparison with particle-tracking experiments. Tracked small particles in the endoplasm should not automatically be interpreted as isolated organelles or molecular cargoes; depending on the preparation and imaging conditions, they may include fragments or local structures of the endoplasm itself. A careful connection between model particles, experimentally tracked objects, and the physical state of the endoplasm will be necessary before quantitative comparison with \textit{Nitella} streaming can be attempted.

\subsection{Limitations and future directions}

The present study intentionally focuses on global averaged quantities: mean transport speed, motor-bound fraction, load, and motor force. This provides a compact characterization of the crossover, but it does not resolve the spatial distribution of dissipation or local force transmission. Future work should examine local velocity fields, spatially heterogeneous resistance, and intermittent transport events, which will be necessary for connecting the present minimal model to richer phenomena such as stop-and-go motion.

The parameter $\gamma_{\mathrm{Slip}}$ should be interpreted as an effective control parameter. Mapping it to a specific microscopic material property would require additional calibration. Similarly, the present parameter scan provides a controlled demonstration of the crossover, while broader parameter scans, additional seeds near the operational crossover region, and system-size checks would be useful for testing robustness.

Overall, the results show that a minimal active transport network can lose motor-activity-dependent transport through increased effective dissipative resistance. This provides a compact physical picture of how an active system may remain mechanically active while crossing over from motor-dominated to drag-dominated transport.

\section{Conclusion}

A minimal active transport network inspired by cytoplasmic streaming was examined to clarify how effective dissipative resistance modifies motor-driven transport. Increasing $\gamma_{\mathrm{Slip}}$ suppressed the mean transport speed without substantially reducing the motor-bound fraction. The mean load and motor force remained finite, indicating that motors remained mechanically active even when net transport was strongly reduced.

The dependence of transport speed on $\alpham$ progressively disappeared as $\gamma_{\mathrm{Slip}}$ increased. This behavior was quantified by the motor dominance ratio and by the $\alpham$-dependent velocity difference, both of which indicate a crossover from a motor-dominated regime to a drag-dominated regime. The crossover is interpreted operationally as a dynamical change in transport response, not as a critical transition.

These results provide a compact physical scenario in which active force generation persists while its contribution to macroscopic transport is suppressed by increased effective dissipative resistance. The model therefore offers a minimal starting point for future studies of how motor activity, mechanical coupling, and dissipation shape cytoplasmic-streaming-inspired transport.

\begin{acknowledgments}
The author thanks researchers at meetings of the Botanical Society of Japan for helpful comments and suggestions. This work was motivated by the long experimental tradition of cytoplasmic-streaming research. Computational analyses were performed using a local multi-node GPU workstation cluster.
\end{acknowledgments}

\bibliographystyle{apsrev4-2}
\bibliography{main}

\end{document}